\documentstyle[twocolumn,epsf]{article}
\newcommand{\beq}{\begin{equation}}
\newcommand{\eeq}{\end{equation}}
\newcommand{\bea}{\begin{eqnarray}}
\newcommand{\eea}{\end{eqnarray}}

\def\nuc#1#2{\relax\ifmmode{}^{#1}{\protect\text{#2}}\else${}^{#1}$#2\fi}

\title{\large
CEM2K - RECENT DEVELOPMENTS IN CEM
}
\vspace{1cm}
\author{Stepan G.~Mashnik and Arnold J.~Sierk\\
{\it Theoretical Division,}
{\it Los Alamos National Laboratory, Los Alamos, NM 87545, USA}} 
\begin{document}
\maketitle
\noindent ABSTRACT
\vspace{0.1cm}

Recent developments of the Cascade-Exciton Model (CEM) of nuclear
reactions are briefly described. 
These changes are motivated by new data on isotope
production measured recently in ``reverse kinematics"  at GSI
for interactions of $^{208}$Pb and $^{238}$U at 1 GeV/nucleon and
$^{197}$Au at 800 MeV/nucleon with liquid $^1$H.
This study leads us to CEM2k, which is a new version of the CEM
code that is still under development.
The increased accuracy and predictive power of 
the code CEM2k are shown by several examples.  Further necessary work is 
outlined.

\vspace{0.4cm}
\noindent INTRODUCTION
\vspace{0.1cm}

The design of a hybrid reactor system driven with a high current accelerator 
requires information about residual nuclides that are produced by high 
energy protons interacting in the target and in structural materials.
It is both physically and economically impossible to measure all necessary
data, which is why reliable models and codes are needed. A model with a 
good predictive power both for the spectra of emitted particles and 
residual nuclide yields is the Cascade-Exciton Model (CEM) of nuclear
reactions [1]. An improved version of the CEM is contained in the code 
CEM95 [2], which is available free from the NEA/OECD, Paris. 
Following an increased interest in intermediate-energy nuclear data in
relation to such projects as Accelerator Transmutation of Wastes (ATW),
Accelerator Production of Tritium (APT), and others,
we developed a new version of the cascade-exciton model, CEM97 [3,4].
CEM97 has a number of improved physics features, uses better 
elementary-particle cross sections for the cascade model, and due to some
significant algorithmic improvements is several
times faster than CEM95. It has been incorporated
into the recent transport code system MCNPX [5].

The recent GSI measurements performed using inverse kinematics
for interactions of $^{208}$Pb [6,7] and $^{238}$U [8] at 1 GeV/nucleon and
$^{197}$Au at 800 MeV/nucleon [9] with liquid $^1$H provide
a very rich set of cross sections for production of practically
all possible isotopes from these reactions in a ``pure" form,
i.e., individual cross sections from a specific given bombarding isotope
(or target isotope, when considering reactions in the usual kinematics,
p + A). Such cross sections are much easier to compare to models than the 
``camouflaged" data from $\gamma$-spectrometry measurements. These 
are often obtained only for a natural composition of isotopes in a target
and are mainly for cumulative production, where measured cross sections
contain contributions nor only from a direct production of
a given isotope, but also from all its decay-chain precursors. 
Analysis of these new data has motivated us to further
improve the CEM and to develop a preliminary version of a new code, 
(called CEM2k), which we describe below.

\vspace{0.4cm}
\noindent RESULTS
\vspace{0.1cm}

First, we discuss briefly the basis of the CEM
and the main differences between 
the improved cascade-exciton model code CEM97 [3,4] and
its precursor, CEM95 [2].
The CEM assumes that the reactions occur in three stages. The first
stage is the IntraNuclear Cascade (INC) 
in which primary particles can be rescattered and produce secondary
particles several times prior to absorption by, or escape from the nucleus.
The excited residual nucleus remaining after the emission of the
cascade particles determines the particle-hole configuration that is
the starting point for the second, pre-equilibrium stage of the
reaction. The subsequent relaxation of the nuclear excitation is
treated in terms of the modified exciton model of pre-equilibrium decay 
followed by the equilibrium evaporative third stage of the reaction.
Generally, all three components may contribute to 
experimentally measured particle spectra and distributions. 

An important ingredient of the CEM is the criterion for transition 
from the intranuclear cascade to the pre-equilibrium model. In 
conventional cascade-evaporation models
(like ISABEL and Bertini's INC used in LAHET [10]) 
fast particles are traced down to some minimal energy, the cutoff energy
$T_{cut}$ (or one compares the duration of the cascade stage of a reaction 
with a cutoff time, in the ``timelike" INC models, like the Liege INC [11])
which is usually about 7--10 MeV above the interior nuclear potential,
below which particles are considered to be absorbed by the
nucleus. The CEM uses a different criterion to decide when a primary
particle is considered to have left the cascade.

An effective local optical absorptive potential $W_{opt. mod.}(r)$ is 
defined from the local interaction cross section of the particle,
including Pauli-blocking effects. This imaginary potential is compared
to one defined by a phenomenological global optical model
$W_{opt. exp.}(r)$. We characterize the degree of similarity or difference
of these imaginary potentials by the parameter 
\begin{equation}
{\cal P} =\mid (W_{opt. mod.}-W_{opt. exp.}) / W_{opt. exp.} \mid .
\end{equation}

When $\cal P$ 
increases 
above
an empirically chosen value, the particle
leaves the cascade, and is then considered to be an exciton.
Both CEM95 and CEM97 use the fixed value $\cal P$ = 0.3.
With this value, we find the cascade stage of the CEM is generally shorter 
than that in other cascade models.

The transition from the preequilibrium stage of a reaction to the
third (evaporation) stage occurs when the probability of nuclear
transitions changing the number of excitons $n$ with
$\Delta n = + 2$ becomes equal to the probability of
transitions in the opposite direction, with $\Delta n = - 2$,
{\it i.e.}, when the exciton model predicts an equilibration has been
established in the nucleus.

The improved cascade-exciton model as
realized in the code CEM97 differs from the CEM95 version by
incorporating new and better approximations for the elementary cross
sections used in the cascade, 
using more precise values for nuclear masses, $Q$-values, binding and 
pairing energies, 
using corrected systematics for the level-density
parameters, 
improving the approximation for the pion ``binding energy", $V_{\pi}$,
adjusting the cross sections of pion absorption on quasideuteron pairs 
inside a nucleus,
considering the effects on cascade particles of refractions and 
reflections from the nuclear potential,
allowing for nuclear transparency of pions,
and including the Pauli principle in the pre-equilibrium calculation.
We also make a number of refinements in the calculation of the
fission channel, described separately in [4]. In addition, we have
improved many algorithms used in the Monte Carlo simulations
in many subroutines, decreasing the computing time by up to a
factor of 6 for heavy targets, which is very important when performing
practical simulations with big transport codes like MCNPX.

The authors of the GSI measurements performed a comparison of their data
to several codes, including LAHET [10], YIELDX [12], ISABELA (see
details and references in [8]), and the Liege INC by Cugnon [11], and 
encountered
serious problems: none of these codes were able to accurately describe 
their measurements; most of the calculated distributions of isotopes 
produced as a function of neutron number were shifted in the direction
of larger masses as compared to the data.
While
in some disagreement with the measurements, the Liege INC and ISABELA
codes provide a better agreement with the data than LAHET and
YIELDX do. Being aware of this situation with the GSI data, 
we decided to consider them ourselves, leading to the development 
of CEM2k.

First, we calculated the $^{208}$Pb GSI data [6] 
with the standard versions
of CEM95 and CEM97. As an example, Fig.~1 shows our first results
obtained with the CEM95 code. Let us note that
so far all CEMxx codes simulate spallation only and do not
calculate any processes following fission, such as production of
fission fragments and the possible evaporation of
particles from them. When, during a simulation of the
compound stage of a reaction these codes encounter a
fission event, they simply tabulate it (allowing
us to calculate the fission cross section and the fissility)
and finish the calculation of this event 
without a subsequent calculation of fission fragments.
Therefore, results from CEM95 shown in Fig.~1
reflect the contribution to the total yields of the 
nuclides only from deep spallation processes 
(successive emission of particles from the target), 
but do not contain fission products.
This is explicitly reflected in smaller calculated cross sections
for light nuclides that are produced partially or mainly by fission.
We will not discuss them here.
To be able to describe 
nuclide production in the fission region, these codes have to be extended
by incorporating a model of high energy fission
({\it e.g.}, in the transport code MCNPX [5], where CEM97 is used, 
it is supplemented by the RAL fission model [13]).

One can see that though CEM95 describes quite well production of 
several heavy isotopes near the target (we calculate p + $^{208}$Pb; 
therefore, for us
$^{208}$Pb is a target not a projectile as in the GSI measurements),
it does not reproduce correctly the cross sections for lighter
isotopes in the deep spallation region. The disagreement
increases with increasing distance from the target, and all calculated
curves are shifted to the heavy mass direction, just as was obtained
by the authors of GSI measurements with all the codes they used.

\begin{center}
\begin{figure*}
\vspace*{-2cm}
\hspace*{-1.5cm}
\includegraphics[angle=-90,width=20.0cm]{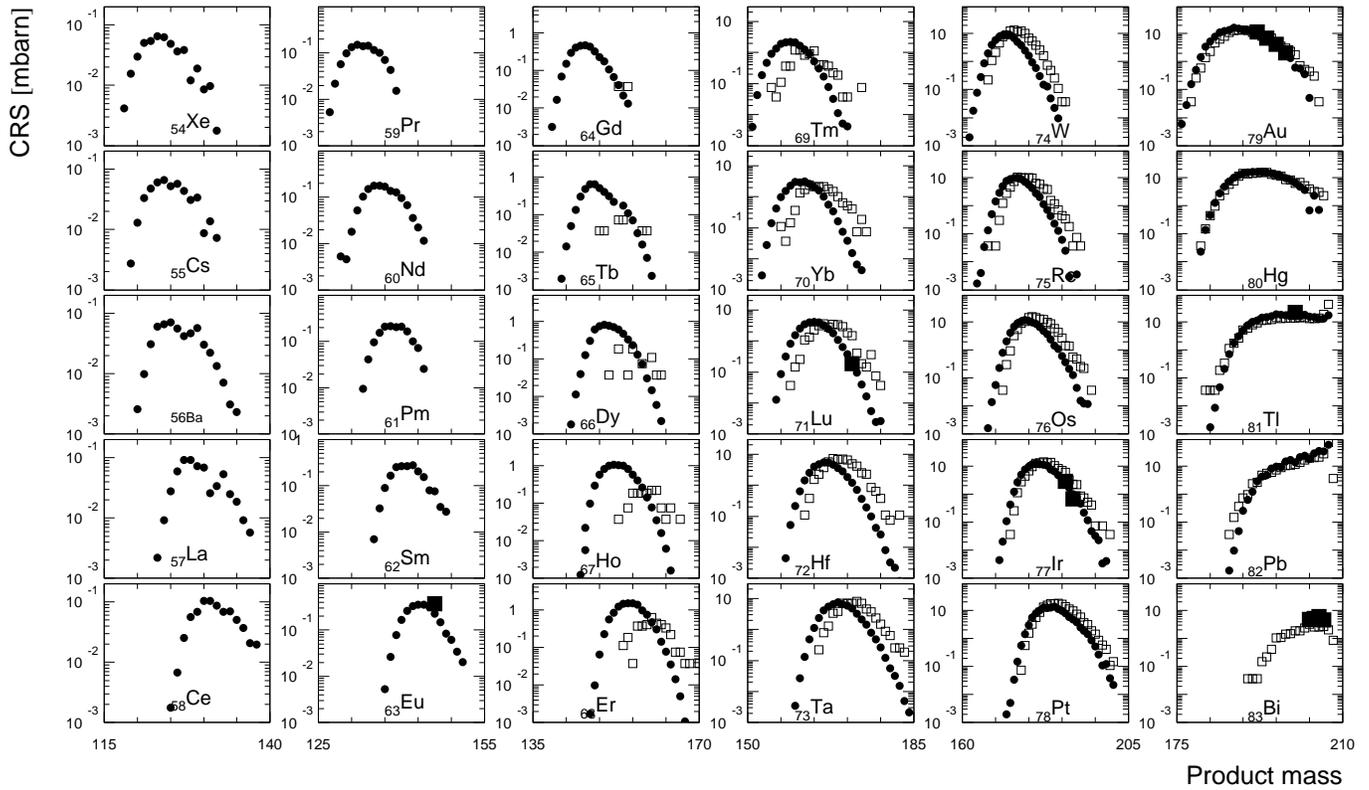}
\vspace*{-2cm}
\caption{Comparison of isotope production cross sections
from $^{208}$Pb + p interaction at 1 GeV/nucleon
measured at GSI [6,7] (filled circles) with the CEM95 calculations
(open squares).
}
\end{figure*}
\end{center}

The results of the CEM97 code are very similar to those of CEM95 shown
in Fig.~1. Our collaborator, Dr.~V.~F.~Batyaev of ITEP, Moscow,
performed an extensive set of calculations of the same data
with several more codes
(HETC, LAHET with both ISABEL and Bertini options, 
CASCADE, CASCADE/INPE, INUCL, and YIELDX) and got very
similar results [14]:
all codes disagree with the data in the deep spallation region,
the disagreement 
increases as one moves away from the target, and all calculated
curves are shifted in the heavy-mass direction.
  
Similar disagreements with the GSI data obtained by different 
authors using different codes at different laboratories suggested 
to us that some general features of these reactions may be 
treated wrong by all codes and we started 
to modify our CEM97 code to attempt to 
understand the source of the disagreement.

Fig.~2 shows some examples of steps we took in trying to determine
which physical parameters might affect the discrepancies.
First, according to Prokofiev's systematics [15], the value of
the fission cross section for the reaction p (1 GeV) + $^{208}$Pb
should be $\approx 100$ mb, i.e., only about 6\% of the total reaction
cross section of 1600 mb, as used by CEM97. In the past,
if we were interested in describing only emitted particles and
yields of the spallation products,
we would not take into account fission processes
in calculations of reactions with such small fission cross sections, 
to make the calculations faster. 
We would never be able predict cross sections with an accuracy better
than 6\%. However, since many of the isotopes are produced with quite
small cross sections, there is the possibility that the variation
of fission barriers with neutron and proton numbers might lead to
an isotopic fractionation effect.
Calculations ignoring fission are shown both in Fig.~1 for all 
isotopes measured at GSI [6] (CEM95) and in Fig.~2a,
for production of Tl, Ir, and Tm, chosen as examples (CEM95 and CEM97).
As the general agreement of calculations with the data is not good,
we decided to look at the influence of fission on the
production of isotopes in the spallation region.
In Fig.~2b, the 
solid lines show calculations with CEM97 including fission,
while dashed lines show the results from CEM97 (from
Fig.~2a) not including fission. We see that though for most of
the isotopes with high yields the two calculations almost coincide,
the difference between the two calculations 
is somewhat larger for production of neutron-deficient 
isotopes of Tl (and to a lesser extent, Ir). We obtain similar
results for other isotopes.
This difference is large enough to lead us to the conclusion
that if we hope to describe correctly production
of all isotopes from a nuclear reaction, we have to take into account 
fission processes, even when the fission cross section itself is quite small.
All the following results calculated by CEM97 are done including
the fission mode, though this will not be mentioned explicitly 
in the text again.


\begin{center}
\begin{figure*}
\vspace*{-3cm}
\centerline{\epsfxsize 21.5cm \epsffile{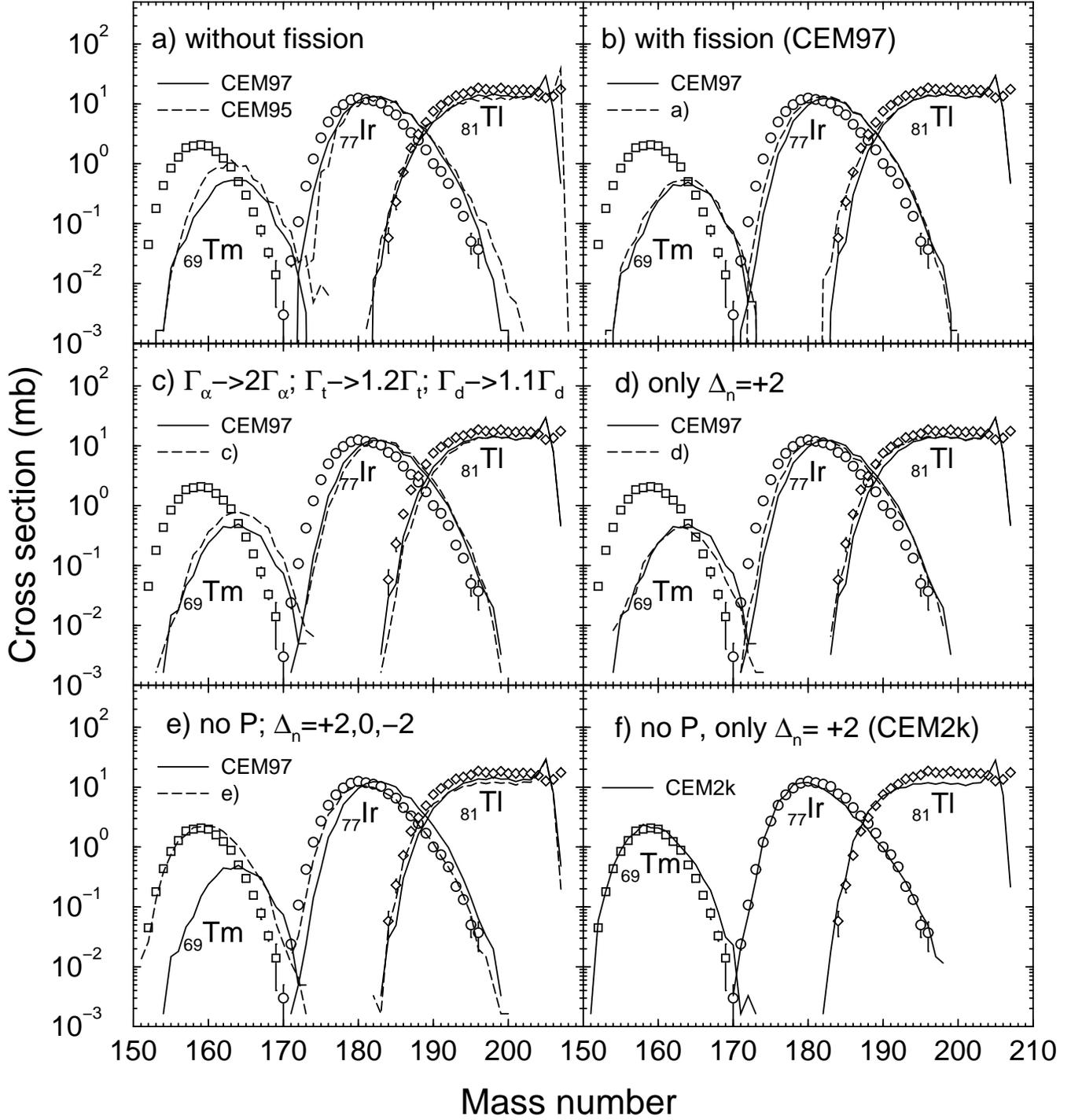}}
\vspace*{-5cm}
\caption{Example of several
steps from CEM97 to CEM2k in analyzing production
of $^{81}$Tl,  $^{77}$Ir, and $^{69}$Tm isotopes from $^{208}$Pb 
interactions with $^1$H
at 1 GeV/nucleon measured at GSI [3] (see details in the
text).
}
\end{figure*}
\end{center}

We know that a currently unsolved problem of CEM97, CEM95,
and, to the best of our knowledge, of most of other available codes,
is the correct prediction of the complex particle yields.
Usually, all codes underestimate production of $^4$He, and to
some extent, also of t and d at these incident energies.
Such underestimation of complex-particle emission could affect
the production of the final residual isotopes, so we decided
to investigate the magnitude of this effect for our case. We are working
now on better inverse cross sections used in calculating the probability
for emission of complex particles and other improvements to the CEM
hoping to address the problem of complex-particle
yields, but for the moment we will
estimate the effect of complex particle emission on residual
nuclide production with a simple approximation.
Our experience from analyzing 
other reactions involving similar energies and targets
tells us that usually we underestimate the multiplicities of
$^4$He by a factor of 2, of t by a factor of 1.2, and of d
by a factor of 1.1. In our model, these particles are produced
only at the preequilibrium and evaporation stages of a reaction 
and their emission is governed by their widths. So, we
multiply ``by hand" the calculated widths of $^4$He by 2,
of t by 1.2, and of d by 1.1 to estimate the effect. Our
results are shown in Fig.~2c) as dashed curves
compared with standard CEM97 results from Fig.~2b shown
as solid curves. Surprisingly, the two sets of calculations
differ by little, so a simple enhancement of complex particle
emission won't solve the problem of most residual nuclide yields.
(Of course, for a specific isotope where $\alpha$ emission
is an important mode of its production, the effect IS significant.)

The standard CEM97 predicts the following multiplicities
for n, p, d, t, $^3$He, and $^4$He: 
$9.2409 \pm 0.0025$,
$1.9624 \pm 0.0011$,
$0.5518 \pm 0.0006$,
$0.1593 \pm 0.0003$,
$0.0682 \pm 0.0002$,
and
$0.1743 \pm 0.0003$,
while if we enlarge the widths of $^4$He, t, and d as described above
we get, respectively,
$9.0886 \pm 0.0025$,
$1.9198 \pm 0.0011$,
$0.5693 \pm 0.0006$,
$0.1714 \pm 0.0003$,
$0.0629 \pm 0.0002$,
  and
$0.02743 \pm 0.0004$.
In other words,
enlarging the widths of $^4$He, t, and d does indeed predict more
$^4$He, t, and d as it should, but at the same time fewer n and p,
so that the net effect on the residual nuclides is minimal. 
We conclude that although the problem of complex-particle yields
needs to be solved independently of the GSI data, solving this problem
will not affect significantly the production of most of the final 
residual nuclides, so we can try to find ways of describing the GSI data
without first solving the problem of complex-particle yields.

One of the things we previously noted was the fact that the calculated
curves for almost all isotopes 
are shifted in the heavy mass direction and the shift
increases as the atomic number decreases from that of the target.
In other words this means that for a given final isotope (Z), all
models predict emission of too few neutrons. Most of the neutrons are
emitted at the evaporation stage of a reaction. So, one way to 
increase the number of emitted neutrons would be to increase the 
evaporative part of a reaction. In our approach, this might be done 
in two different ways: the first would be to have a shorter 
preequilibrium stage, so that more excitation energy remains available
for the following evaporation; the second would be to have a longer
cascade stage of a reaction, so that after the cascade, less exciton
energy is available for the preequilibrium stage, so fewer energetic
preequilibrium particles are emitted, leaving more excitation energy
for the evaporative stage.

Preequilibrium decay in our model involves both emission of particles
from particle-like quasiparticle states, and the evolution of the
number of excitons via equations for changes in exciton number of
$\Delta  n = + 2, 0, - 2$ [1].
We make the transition to evaporation when the number of excitons
exceeds a critical number which depends on the nuclear mass and 
excitation energy.
One easy way to shorten the preequilibrium stage of a reaction
in CEM is to take into account only transitions 
that increase the number of excitons, 
$\Delta  n = + 2$, i.e.,
the evolution of a nucleus only torward the compound nucleus,
during the equilibration. We would thus not take into account
possible transitions backward, decreasing the number of excitons 
$\Delta n = - 2$ 
and transitions without changing the number of excitons
$\Delta n = 0$. In this case, the time of the
equilibration will be shorter and fewer
preequilibrium particles will be emitted, leaving more excitation
energy for the evaporation. Such an approach is used by some exciton
models, for instance, by the Multistage Pre-equilibrium Model
used in LAHET [10].
Calculations in this modification to CEM97 are shown in
Fig.~2d by dashed curves, compared with the standard CEM97 results
shown by solid curves. We see that we get a slight improvement
and a shift of the calculated curves in the right direction, but the
effect is too small and doesn't solve the problem.

In Fig.~2e, we show 
the results of the second 
method of increasing evaporation.
To enlarge the cascade part of a reaction in CEM we
have either to enlarge the parameter $\cal P$ or to remove it completely
and use a cutoff energy $T_{cut}$, as do other INC models.
Our calculations have shown that a reasonable increase 
of $\cal P$ doesn't solve the problem.
The dashed curves in Fig.~2e show calculations with CEM97 not 
using the parameter $\cal P$, instead using a cutoff energy 
of $T_{cut} = 1$ MeV, again compared with our standard CEM97 results
shown by the solid curves. One can see a significant improvement
of the agreement with the GSI data in this approach. Only a little
more neutron emission is required to get just
a perfect agreement with the data.  We choose to achieve this
by applying in addition to this approach the
condition of taking into account only transitions with
$\Delta n = +2$, discussed above in reference to Fig.~2d. Using both these
conditions leads to the results shown in Fig.~2f.
We call this approach CEM2k and see that it
describes the GSI data very well.

The behavior of a compound nucleus in CEM is governed by
its mass number, A, charge, Z, excitation energy, $E^*$, and its
angular momentum, L (we do not have in CEM any information about
the deformation of a nucleus after the preequilibrium stage
of a reaction). It is informative to look at these quantities 
for the compound nuclei remaining after the preequilibrium stage
for the different approaches discussed above. Table 1 shows
values of the mean A, Z, $E^*$ and L of the compound nuclei calculated
using all the  discussed approaches.
One can see that as a result of the transition from CEM97 (plots a and b)
to CEM2k (plot f), the mean mass and charge of the compound nuclei
are almost the same, while their mean excitation energy increases
by about 40 MeV. This is the main factor leading to a larger
number of evaporated neutrons, and as a consequence,
to a much better agreement with the GSI data.

\begin{table}[ht]\caption{Mean mass number A, charge Z, excitation  
energy $E^*$ (MeV), and angular momentum L ($\hbar$) of the
compound nuclei formed after the preequilibrium stage
of reactions calculated by the different approaches as plotted in Fig.~2}
\vspace{0.2cm}
\begin{tabular}{|p{2.4cm}|p{1.0cm}|p{1.0cm}|p{1.0cm}|p{0.8cm}|}\hline
Method& A & Z & $E^*$ & L \\  \hline
CEM97  & 193.9 & 78.2 & 58.7 & 24.6 \\
(plots a and b)& & & & \\ \hline
$\Gamma _{\alpha} \to 2 \Gamma _{\alpha}$ & 193.4 & 78.0 & 58.3 & 24.8 \\
(plot c)& & & & \\ \hline
Only $\Delta _n = + 2 $ & 196.4 & 78.6 & 86.4 & 24.1 \\
(plot d)& & & & \\ \hline
No $\cal P$  & 191.0 & 77.5 & 65.0 & 21.3 \\
(plot e)& & & & \\ \hline
CEM2k $ $ & 193.9 & 78.1 & 97.5 & 20.8 \\
(plot f)& & & & \\       
\hline\end{tabular}
\vspace{0.4cm}
\end{table}

In short, CEM2k has a longer cascade stage,
less preequilibrium emission, and a longer evaporation stage
with a higher excitation energy, as compared to CEM97 and CEM95.

Fig.~3 shows a comparison of the CEM2k results with all 
the data (excluding those involving fission-fragment production)
on isotope yields measured at GSI for the reaction $^{208}$Pb + p.
There is a very good agreement for almost all isotopes.

\begin{center}
\begin{figure*} 
\hspace*{-1.5cm}
\includegraphics[angle=-90,width=20.0cm]{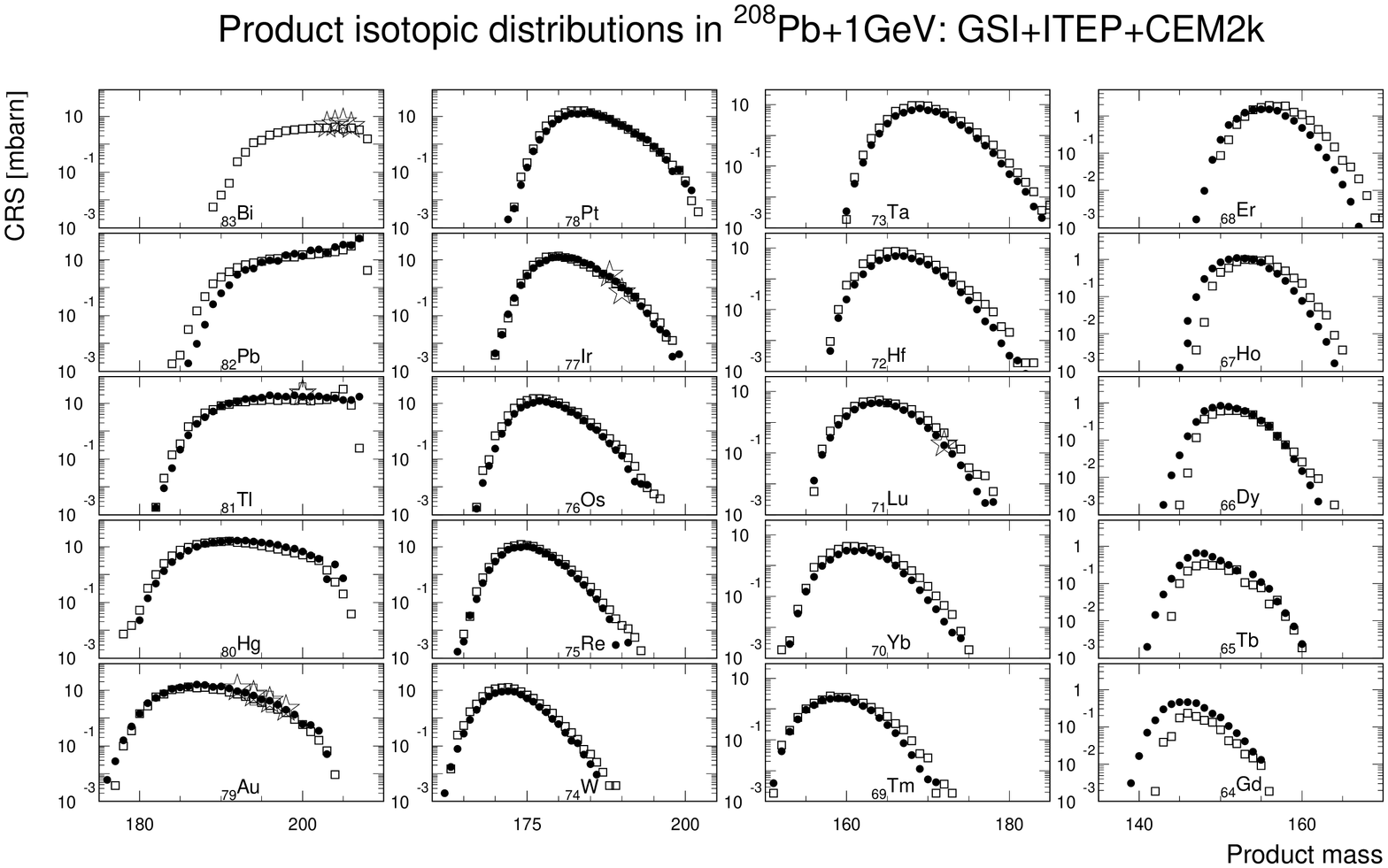}
\vspace*{-20mm}
\caption{Isotopic production cross sections of elements between
Z=64 and 83 in the reaction of $^{208}$Pb on hydrogen at 1 GeV/nucleon.
Filled circles show GSI data [6], open stars are recent ITEP
data measured by the $\gamma$-spectrometry method [16],
while open squares show our CEM2k results.
}
\label{fig6b}
\end{figure*}
\end{center}

A comparison of the CEM2k results with predictions by CEM95 [2],
LAHET-ISABEL [10],
LAHET-Bertini [10], CASCADE/INPE [17], CASCADE [18], YIELDX [12],
and INUCL [19] codes
together with the GSI [6] and ITEP [16] data from our recent paper
[16] is shown in Fig.~4. One can see that CEM2k agrees best with the
data of the other codes tested here.

\begin{center}
\begin{figure*} 
\includegraphics[angle=-90,width=17.7cm]{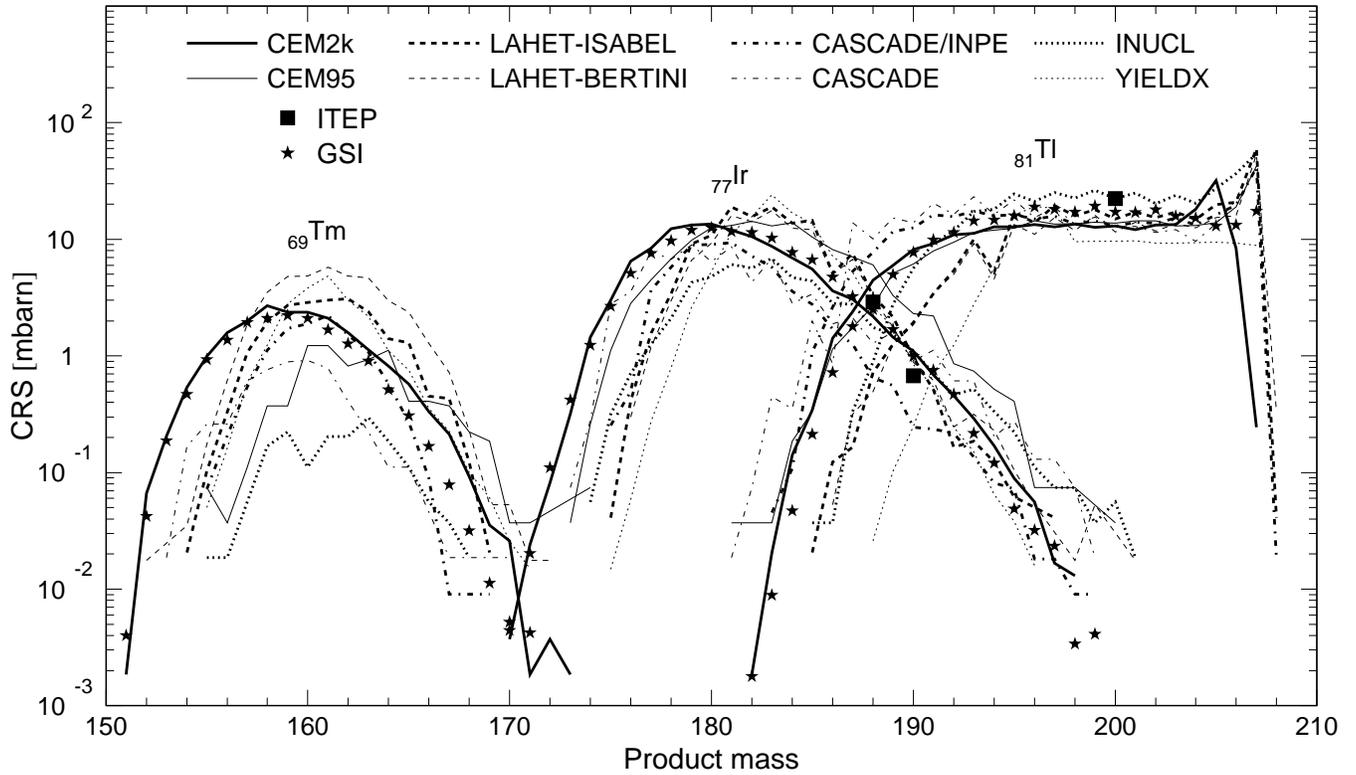}
\vspace*{2mm}
\caption{Isotopic mass distributions for independent production
of Tm, Ir, and Tl isotopes from 1 GeV protons colliding with $^{208}$Pb.
Squares are ITEP measurements [16],
while stars show GSI data obtained in reverse kinematics [6].
Results from different codes are marked as:
CEM2k---our results, CEM95---Ref.~[2], LAHET-ISABEL---Ref.~[10], 
LAHET-Bertini---Ref.~[10], CASCADE/INPE---Ref.~[17], CASCADE---Ref.~[18],
INUCL---Ref.~[19], and YIELDX---Ref.~[12].
}
\label{fig6b}
\end{figure*}
\end{center}
  
Finding a good agreement of CEM2k with the isotope
production yields, we wish to see how well it describes
spectra of secondary particles in comparison with its precursor,
CEM97. Fig.~4 shows examples of neutron spectra from interactions
of protons with the same target, $^{208}$Pb at 0.8 Gev and 1.5 GeV
(we do not know of measurements of spectra at 1 GeV, 
the energy of isotope-production yield data). We see that
CEM2k describes spectra of secondary neutrons certainly no worse than
does CEM97, but possibly a little better at larger angles, in a 
good agreement with the data.
So this preliminary version of an improved CEM code, CEM2k,
describes both the GSI data from $^{208}$Pb interactions with p
at 1 GeV/nucleon and the spectra of emitted neutrons from p+$^{208}$Pb
at 0.8 and 1.5 GeV better than its precursor CEM97.

\begin{center}
\begin{figure} 
\hspace*{-2.3cm}
\includegraphics[angle=0,width=13.5cm]{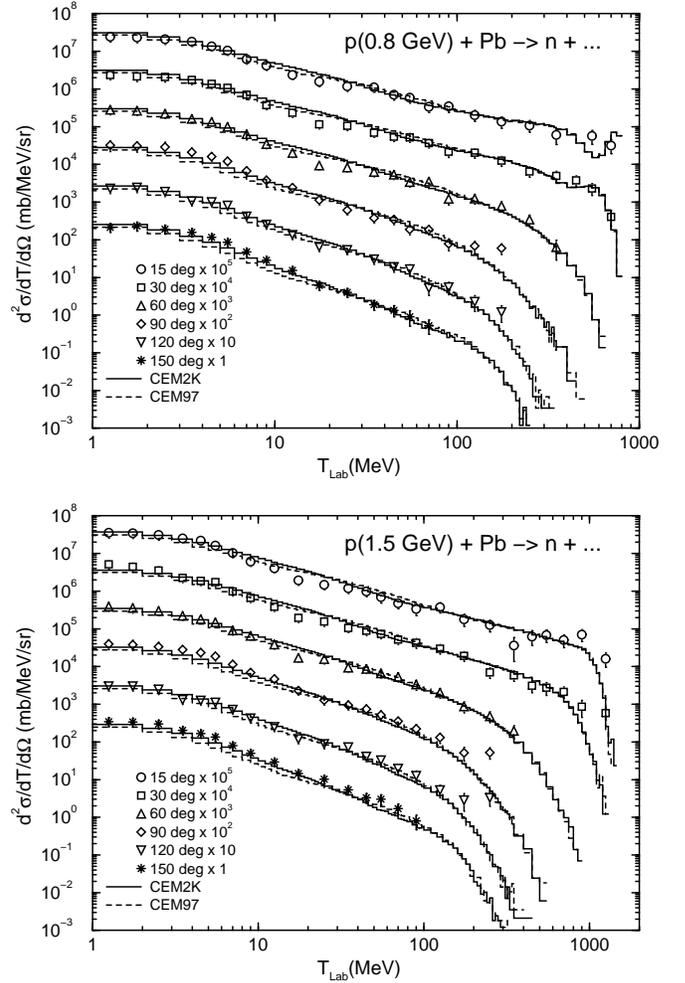}
\vspace*{-25mm}
\caption{Comparison of measured [20] double differential cross
sections of neutrons from 0.8 and 1.5 GeV protons on Pb with
CEM2k and CEM97 calculations.
}
\label{fig6b}
\end{figure}
\end{center}

Besides the $^{208}$Pb data discussed above, 
reactions induced by $^{197}$Au at 800 MeV/nucleon [9]
and $^{238}$U at 1 GeV/nucleon [8] were measured recently at GSI. 
We use CEM2k as fixed from our analysis of the
$^{208}$Pb data [6,7] without further modifications
to calculate the $^{197}$Au measurements [9].
Our CEM2k results are shown in Figs.~6 and 7 together with standard
CEM97 predictions and calculations by LAHET-Bertini [10] and YIELDX [12]
codes from [9]. We see that just as in the case of the $^{208}$Pb data,
CEM2k agrees best with the $^{197}$Au data in the spallation region
compared to the other codes tested here. For the production of nuclides
lighter than Tb (Fig.~7), where fission-fragment formation begins
to contribute, CEM2k underestimates the data (it doesn't include
a model of fission-fragment production, as we discussed previously).

\begin{center}
\begin{figure*} 
\vspace*{-50mm}
\hspace*{-3.0cm}
\includegraphics[angle=-0,width=24.0cm]{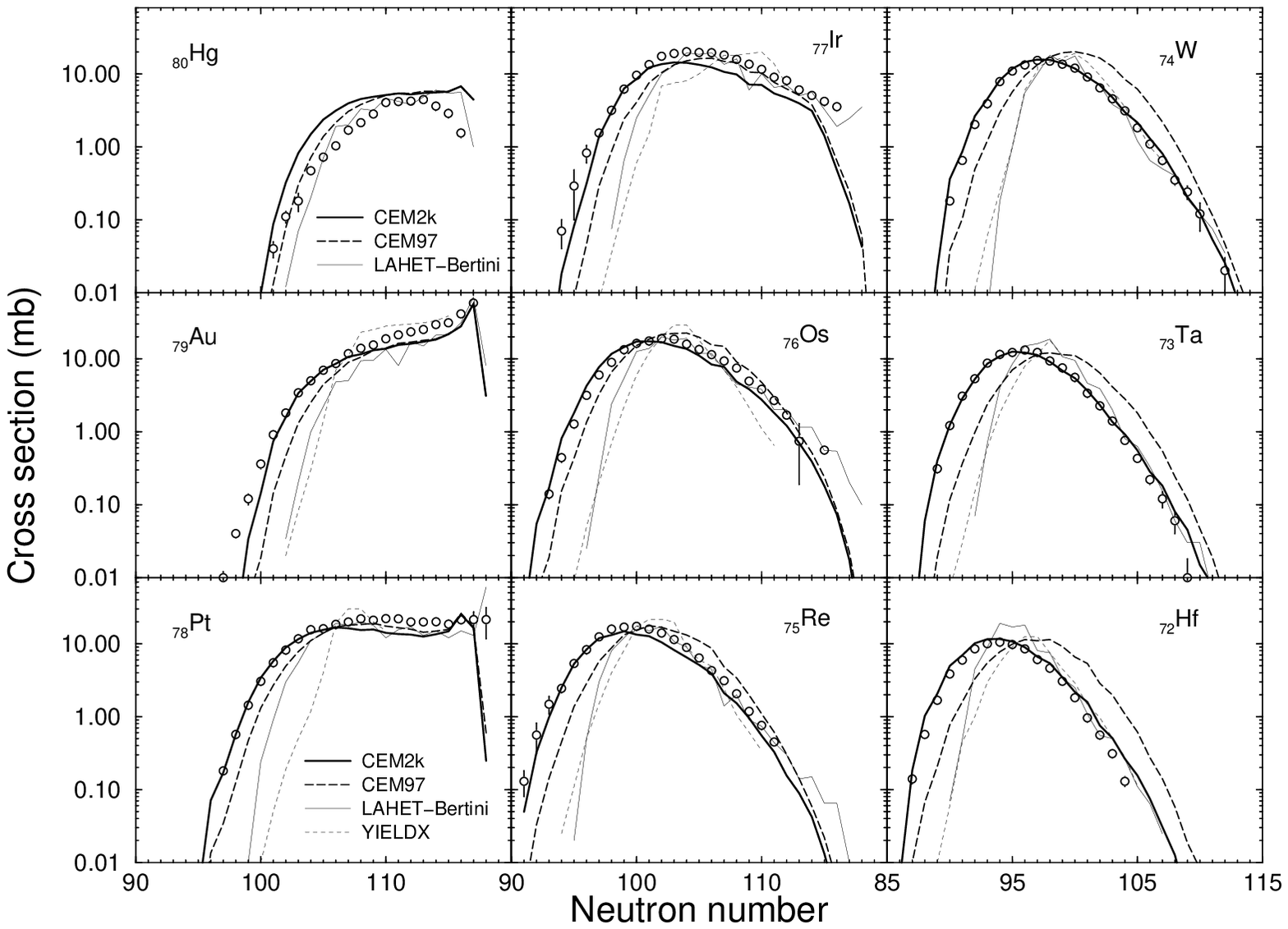}
\vspace*{-150mm}
\caption{Isotopic distribution of the spallation products from 
the reaction $^{197}$Au + p at 800 $A$ MeV from mercury to hafnium.
Open circles are the GSI data [9], CEM2k (thick solid curves)
and CEM97 (thick dashed curves) are our present calculations,
LAHET-Bertini (thin solid curves) and YIELDX (thin dashed curves)
are results of calculations from [9].
}
\end{figure*}
\end{center}

\begin{center}
\begin{figure*} 
\vspace*{-60mm}
\hspace*{-3.0cm}
\includegraphics[angle=-0,width=24.0cm]{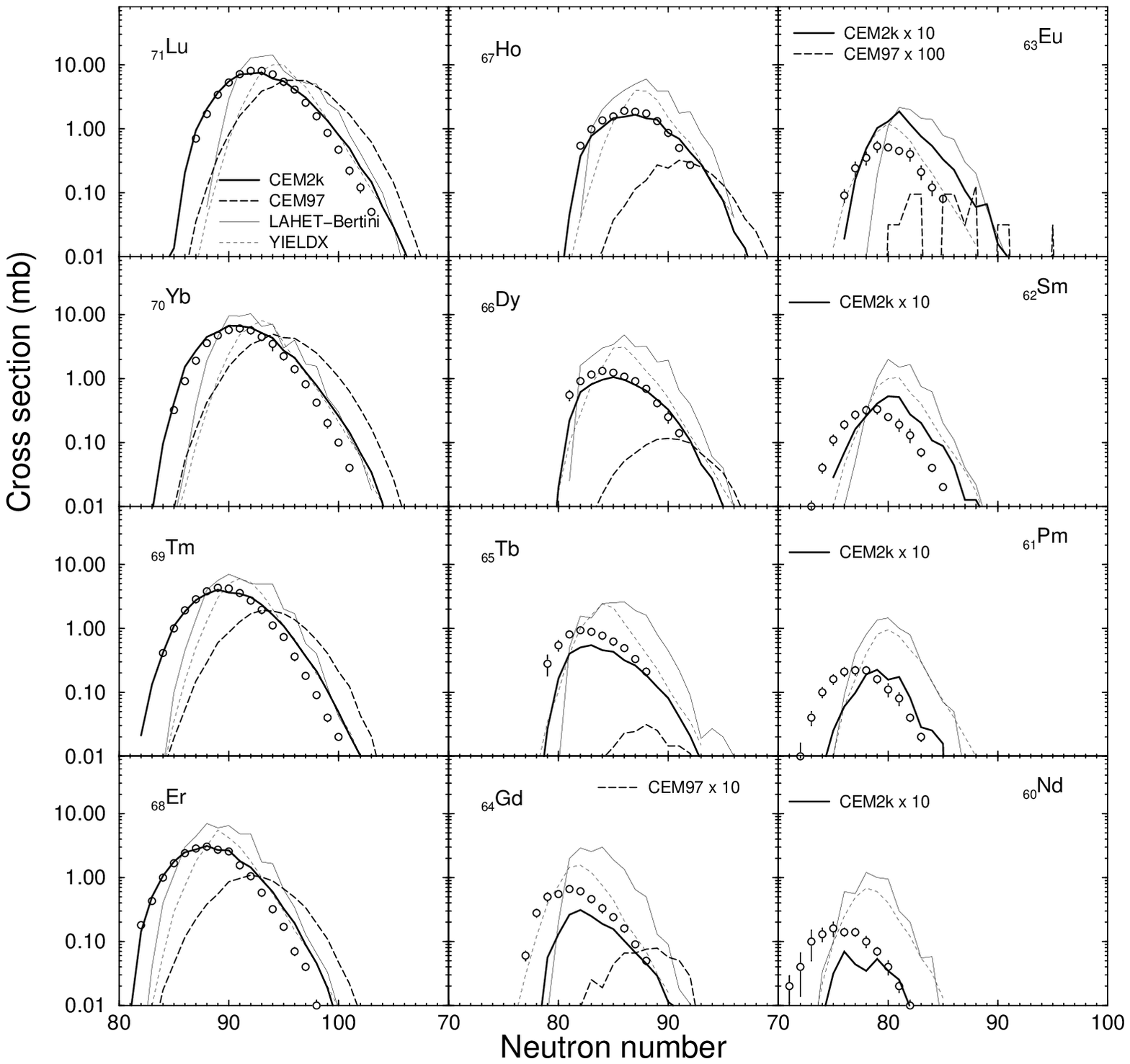}
\vspace*{-110mm}
\caption{The same as Fig.~6 but for products from
lutetium to neodymium. 
}
\end{figure*}
\end{center}

The $^{238}$U GSI data are still preliminary (with relative 
errors of 40\%), 
published only as figures [8] and not available yet 
in tabulated form. These data are of a great interest to
us as the reaction mechanisms in the case of p + $^{238}$U interactions
differ significantly from those of $^{208}$Pb and $^{197}$Au
targets.
Again, according to Prokofiev's systematics [15],
for the reaction p (1 GeV) + $^{238}$U the fission cross section is
about 1390 mb, which makes up about 80\% of the total reaction 
cross section of 1788 mb as calculated by CEM2k. This means that
fission is the main mode for this reaction and almost everything
in a model prediction of isotope production 
should be determined by how well the model describes the
neutron-to-fission width ratio, $\Gamma_n / \Gamma _f$. So, the
$^{238}$U GSI data will be very useful to test and improve
the treatment of the fission mode by any code, including CEM2k.
We enlarged the figures from [8], to get (approximate!) numerical 
values of measured cross sections, and calculated
this reaction with CEM2k.

As an example, Fig.~8 shows a comparison of our CEM2k calculations
with the GSI data [8] on production of uranium isotopes from
p (1 GeV) interactions with $^{238}$U and of gold isotopes from proton
interactions with $^{197}$Au. The agreement of CEM2k with $^{238}$U data
is not as good as for $^{197}$Au, or as observed above for $^{208}$Pb
at 1 GeV/nucleon and $^{197}$Au at 800 MeV/n. This means we still have
room for CEM2k improvement, especially in the treatment of the
fission mode. At the same time, Fig.~8 demonstrates how different are
the reaction mechanisms in $^{197}$Au and $^{238}$U: while 21 isotopes
of gold were measured from $^{197}$Au, only 10 isotopes of uranium
were measured from $^{238}$U, and their yield decreases exponentially
as one moves into the neutron-deficient region, a direct result
of the dominant role of the fission mode in uranium.

\begin{center}
\begin{figure} 
\hspace*{-1.2cm}
\includegraphics[angle=-90,width=11.0cm]{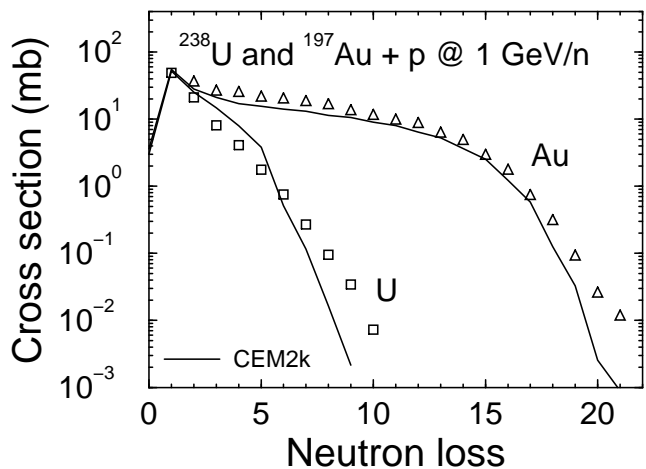}
\vspace*{-10mm}
\caption{Comparison of the production of projectile 
isotopes in gold and uranium interactions with liquid 
hydrogen targets as measured at GSI [8] (symbols)
and predicted by CEM2k (curves).
}
\label{fig6b}
\end{figure}
\end{center}

\vspace*{0.2cm}
\noindent FURTHER WORK
\vspace{0.1cm}

Besides the changes to  CEM97 mentioned above, we also made a 
number of other improvements and refinements, such as 
imposing momentum-energy conservation for each simulated event
(the Monte Carlo algorithm previously used in CEM 
provides momentum-energy conservation only 
statisticaly, on average, but not exactly for each simulated
event);
using real binding energies for nucleons at the cascade 
stage of a reaction instead of the approximation of a constant
separation energy of 7 MeV used in previous versions of the CEM; 
using reduced masses of particles in the calculation their
emission widths instead of using the approximation
of no-recoil used in previous versions;
coalescence of complex particles from fast cascade nucleons
already outside the nucleus.
These refinements are important physically, and they slightly
improve the agreement of calculations with the data,
but not so dramatically as the changes discussed above.
For this reason we do not show here figures with examples.
In addition, CEM2k is still under development and will still change.
We need to analyze a lot of more available data with it,
especially at lower incident energies. We plan also to incorporate
better inverse cross sections and to solve a problem of 
overestimation of fission cross sections at energies above 1 GeV,
observed recently in our preliminary calculations of the p (2.6 GeV) + Hg
reaction [21].
When we complete CEM2k to a reasonable level, we plan
to incorporate it into LAHET and to replace the present CEM97 
in MCNPX and the present CEM95 in the MARS code system [22].

\vspace{0.4cm}
\noindent ACKNOWLEDGMENTS
\vspace{0.1cm}

We thank Dr.~B.~Mustapha for sending us tables with GSI data
for $^{208}$Pb and $^{197}$Au and express our gratitude to
Drs.~S.~Chiba and K.~Ishibashi for suppling us with numerical values
of their neutron spectra measurements. We are grateful to Drs. L.~S.~Waters,
R.~E.~Prael, and N. V. Mokhov for useful discussions, interest, and support.

This study was supported by the U.~S.~Department of Energy.

\vspace{0.5cm}
\noindent REFERENCES
\vspace{0.1cm}
\begin{enumerate}

\vspace*{-0.2cm}
\item 
K.~K.~Gudima, S.~G.~Mashnik, and V.~D.~Toneev,
``Cascade-Exciton Model of Nuclear Reactions,"
{\it Nucl.~Phys.}~{\bf A401}, 329 (1983).

\vspace*{-0.2cm}
\item 
S.~G.~Mashnik,~``User~Manual~for~the Code CEM95," JINR, Dubna (1995);
OECD NEA Data Bank, 
Paris, France (1995); \\
http://www.nea.fr/abs/html/iaea1247.html;
RSIC-PSR-357, Oak Ridge, 1995.

\vspace*{-0.2cm}
\item
S.~G.~Mashnik and A.~J.~Sierk,
``Improved Cascade-Exciton Model of Nuclear Reactions,"
{\em Proc.~Fourth Workshop on Simulating Accelerator Radiation Environments
(SARE4)}, Knoxville, TN, USA, September 14--16, 1998, 
T.~A.~Gabriel, Ed., ORNL (1999) 
pp.~29--51.

\vspace*{-0.2cm}
\item
A.~J.~Sierk and S.~G.~Mashnik,
``Modeling Fission in the Cascade-Exciton Model,"
{\em Proc.~Fourth Workshop on Simulating Accelerator Radiation Environments
(SARE4)}, Knoxville, TN, USA, September 14--16, 1998, 
T.~A.~Gabriel, Ed., ORNL (1999) 
pp.~53--67.

\vspace*{-0.2cm}
\item
{\it MCNPX$^{TM}$ User's Manual, Version 2.1.5}, edited by Laurie S.~Waters,
Los Alamos National Laboratory Report LA-UR-99-6058 (1999).

\vspace*{-0.2cm}
\item
W.~Wlazlo, T.~Enqvist, P.~Armbruster, J.~Benlliure,
M.~Bernas, A.~Boudard, S.~Cz\'ajkowski, R.~Legrain, S.~Leray,
B.~Mustapha, M.~Pravikoff, F.~Rejmund, K.-H.~Schmidt, 
C.~St\'ephan, J.~Taieb, L.~Tassan-Got, and C.~Volant,
``Cross Sections of Spallation Residues Produced in 
1$A$ GeV $^{208}$Pb on Proton Reactions,"
{\em Phys.~Rev.~Lett.}~{\bf 84}, 5736 (2000).

\vspace*{-0.2cm}
\item
T.~Enqvist, W.~Wlazlo, P.~Armbruster, J.~Benlliure,
M.~Bernas, A.~Boudard, S.~Cz\'ajkowski, R.~Legrain, S.~Leray,
B.~Mustapha, M.~Pravikoff, F.~Rejmund, K.-H.~Schmidt, 
C.~St\'ephan, J.~Taieb, L.~Tassan-Got, and C.~Volant,
``Isotopic Yields and Kinematic Energies of Primary Residues in 
1$A$ GeV $^{208}$Pb + p Reactions,"
GSI Preprint 2000-28, 
http://www-wnt.gsi.de/kschmidt/publica.htm,
submitted to {\em Nucl.~Phys.~A}.

\vspace*{-0.2cm}
\item 
J.~Taieb, P.~Armbruster, J.~Benlliure, M.~Bernas, A.~Boudard, 
S.~Cz\'ajkowski, T.~Enqvist, F.~Rejmund, R.~Legrain, S.~Leray, B.~Mustapha, 
M.~Pravikoff, K.-H.~Schmidt, C.~St\'ephan, L.~Tassan-Got, C.~Volant, and
W.~Wlazlo, 
``Measurement of $^{238}$U Spallation Product Cross Sections at 
1 GeV per Nucleon,"
http://www-wnt.gsi.de/kschmidt/publica.htm\#Conferences.

\vspace*{-0.2cm}
\item 
F.~Rejmund, B.~Mustapha, P.~Armbruster, 
J.~Benlliure, M.~Bernas, A.~Boudard, J.~P.~Dufour,
T.~Enqvist, R.~Legrain, S.~Leray, K.-H.~Schmidt, 
C.~St\'ephan, J.~Taieb, L.~Tassan-Got, and C.~Volant,
``Measurement of Isotopic Cross Sections of Spallation Residues 
in 800 $A$ MeV $^{197}$Au + p Collisions,"
GSI Preprint 2000-06, 
http://www-wnt.gsi.de/kschmidt/publica.htm,
submitted to {\em Nucl.~Phys.~A}.

\vspace*{-0.2cm}
\item 
R.~E.~Prael and H.~Lichtenstein, 
``User guide to LCS: The LAHET Code System," 
LANL Report No.~LA-UR-89-3014, Los Alamos (1989); \\
http://www-xdiv.lanl.gov/XTM/lcs/lahet-doc.html.

\vspace*{-0.2cm}
\item 
J.~Cugnon, C.~Volant, and S.~Vuillier,
``Improved Intranuclear Cascade Model for Nucleon-Nucleus Interactions,
{\em Nucl.~Phys.}~{\bf A620}, 475 (1997).

\vspace*{-0.2cm}
\item 
R.~Silberberg, C.~H.~Tsao, and A.~F.~Barghouty,
``Updated Partial Cross Sections of Proton-Nucleus Reactions,"
{\em Astrophys.~J.}~{\bf 501}, 911 (1998);
R.~Silberberg and  C.~H.~Tsao, 
``Partial Cross-Sections in High-Energy Nuclear Reactions, and
Astrophysical Applications. I. Targets With Z $\le 28$,"
{\em Astrophys.~J.~Suppl.}, No.~220, {\bf 25}, 315 (1973);
R.~Silberberg and  C.~H.~Tsao, 
``Partial Cross-Sections in High-Energy Nuclear Reactions, and
Astrophysical Applications. II. Targets Heavier Than Nickel,"
{\em ibid}, p.~335.

\vspace*{-0.2cm}
\item 
F.~Atchison,
``Spallation and Fission in Heavy Metal Nuclei under Medium
Energy Proton Bombardment," 
in {\it Targets for Neutron Beam Spallation Source},
Jul-Conf-34, Kernforschungsanlage Julich GmbH (January 1980).

\vspace*{-0.2cm}
\item 
V.~F.~Batyaev, private communication, June 2000, and to be published.

\vspace*{-0.2cm}
\item 
A.~V.~Prokofiev,
``Compilation and Systematics of Proton-Induced Fission Cross Section Data,"
{\em Nucl.~Instrum.~Meth.~A}, to be published in 2000.

\vspace*{-0.2cm}
\item 
Yu.~E.~Titarenko, O.~V.~Shvedov, V.~F.~Batyaev, E.~I.~Karpikhin,    
V.~M.~Zhivun, A.~B.~Koldobsky, R.~D.~Mulambetov, D.~V.~Fischenko,   
S.~V.~Kvasova, A.~N.~Sosnin,
S.~G.~Mashnik, R.~E.~Prael, A.~J.~Sierk,
T.~A.~Gabriel, M.~Saito, and H.~Yasuda,
``Cross Sections for Nuclide Production in 1 GeV Proton-Irradiated
$^{208}$Pb,"
LANL Report LA-UR-00-4779, Los Alamos (2000),
submitted to Phys.~Rev.~C.

\vspace*{-0.2cm}
\item 
V.~S.~Barashenkov, A.~Yu.~Konobeev, Yu.~A.~Korovin, and V.~N.~Sosnin,
``CASCADE/INPE Code System,"
{\em Atomic Energy}, {\bf 87}, 742 (1999).

\vspace*{-0.2cm}
\item 
V.~S.~Barashenkov, Le Van Ngok, L.~G.~Levchuk, Zh.~Zh.~Musul'manbekov, 
A.~N.~Sosnin, V.~D.~Toneev, and S.~Yu.~Shmakov, 
JINR Report R2-85-173, Dubna, 1985; 
V.~S.~Barashenkov, F.~G.~Zheregi, and Zh. Zh. Musul'manbekov, 
{\it Sov. J. Nucl. Phys.}, {\bf 39}, 715 (1984); 
V. S. Barashenkov, B.~F.~Kostenko, and A.~M.~Zadorogny, 
{\it Nucl.~Phys.},~{\bf A338}, 413 (1980).

\vspace*{-0.2cm}
\item 
G.~A.~Lobov, N.~V.~Stepanov, A.~A.~Sibirtsev, and Yu.~V.~Trebukhovskii, 
Institute for Theoretical and Experimental Physics (ITEP) 
Preprint No.~ITEP-91, Moscow, 1983; 
A.~A.~Sibirtsev, N.~V.~Stepanov, and Yu.~V.~Trebukhovskii, 
ITEP Preprint No.~ITEP-129, Moscow, 1985; 
N.~V.~Stepanov, ITEP Preprint No.~ITEP-81, Moscow, 1987; 
N.~V.~Stepanov,  ITEP Preprint No.~ITEP-55-88, Moscow, 1988 (in Russian).

\vspace*{-0.2cm}
\item 
K.~Ishibashi, 
H.~Takada, T.~Nakomoto, N.~Shigyo, K.~Maehata, N.~Matsufuji,
S.~Meigo, S.~Chiba, M.~Numajiri, Y.~Watanabe, T.~Nakamura,
``Measurement of Neutron-Production Double-Differential
Cross Sections for Nuclear Spallation Reaction Induced by 0.8,
1.5 and 3.0 GeV Protons,"
{\em J.~Nucl.~Sci.~Techn.}, {\bf 34}, 529 (1997).

\vspace*{-0.2cm}
\item 
Yu.~E.~Titarenko, O.~V.~Shvedov, V.~F.~Batyaev, V.~M.~Zhivun,  
E.~I.~Karpikhin,    
R.~D.~Mulambetov, D.~V.~Fischenko,   
S.~V.~Kvasova, ,
S.~G.~Mashnik, R.~E.~Prael, A.~J.~Sierk,
and H.~Yasuda,
``Study of Residual Product Nuclide Yields From 0.1, 0.2, 0.8,
and 2.6 GeV Proton-Irradiated $^{nat}$Hg Targets,"
LANL Report LA-UR-00-3600, Los Alamos (2000);
E-print nucl-ex/0008012 21 Aug 2000;
to be published in 
{\it Proceedings of the Fifth International Workshop on 
Shielding Aspects of Accelerators, Irradiation and Target Facilities
(SATIF5)},
July 18--21, 2000, OECD Headquarters, Paris, France.

\vspace*{-0.2cm}
\item
N. V. Mokhov, S. I. Striganov, A. Van Ginneken, S.~G.~Mashnik,
A.~J.~Sierk, and J. Ranft,
``MARS Code Developments,"
{\em Proc.~Fourth Workshop on Simulating Accelerator Radiation Environments
(SARE4)}, Knoxville, TN, USA, September 14--16, 1998, 
T.~A.~Gabriel, Ed., ORNL (1999) 
pp.~87--99.

\end{enumerate}

\end{document}